\documentclass[aps,prl,twocolumn,showpacs,groupedaddress]{revtex4}

\usepackage{graphicx}
\begin{document}
\title{Off-diagonal geometric phase in composite systems}
\author{X. X. Yi$^{1,2}$, J. L. Chang $^1$}
\affiliation{$^1$Department of physics, Dalian University of
Technology, Dalian 116024 China\\
$^2$ Institute of Theoretical Physics, Chinese Academy of
Sciences, Beijing 100080, China }
\date{\today}
\begin{abstract}
The effect due to the inter-subsystem coupling  on the
off-diagonal geometric phase in a composite system is
investigated. We analyze the case where the   system undergo an
adiabatic evolution. Two coupled qubits driven by time-dependent
external magnetic fields are presented as an example, the
off-diagonal geometric phase as well as the adiabatic condition
are  examined and discussed.
\end{abstract}
 \maketitle

A quantal system in a eigenstate $|\psi_j(\vec{s}\rangle$,
adiabatically transport round a circuit by varying parameters
$\vec{s}=(s_1,s_2,...)$ in its Hamiltonian $H(\vec{s})$, will
acquire a geometric phase $\Gamma_g$ in addition to the familiar
dynamical phase $\Gamma_d$ given by $\Gamma_d+\Gamma_g=
\text{arg}\langle \psi_j(\vec{s}_1)|\psi_j(\vec{s}_2)\rangle$;
$\Gamma_g$ is the well known geometric phase(or Berry phase) when
$\vec{s}_1=\vec{s}_2$ and the state $|\psi_j(\vec{s})\rangle$ is
transported adiabatically along a closed loop \cite{berry}. The
geometric phase has been extensive studied
\cite{shapere,thouless,sun} and generalized to non-adiabatic
evolution \cite{aharonov}, mixed states
\cite{uhlmann86,sjoqvist1,singh}, and open systems \cite{carollo}.
Recent studies on the geometric phases found that the (diagonal)
geometric phase itself could not exhaust all information
containing in phases acquired when the quantal system undergo an
adiabatic evolution, this can be understood as follows. Consider
parallel transport generated by the operator $U^{\parallel}$ of
the $j$th eigenstate $|\psi_j(\vec{s})\rangle$, in the case of
$|\psi_k(\vec{s}_2)\rangle=U^{\parallel}
|\psi_j(\vec{s}_1)\rangle$ $(j\neq k)$ the scalar product $\langle
\psi_j(\vec{s}_2)|U^{\parallel}|\psi_j(\vec{s}_1)\rangle$ vanishes
and the (diagonal) geometric phase becomes undefined. The only
information left is the cross scalar product $\langle
\psi_k(\vec{s}_2)|U^{\parallel}|\psi_j(\vec{s}_1)\rangle$, this
gives rise to the definition of the off-diagonal geometric phase
factor \cite{manini}($j\neq k$)
\begin{eqnarray}
\gamma_{jk}&=&\sigma_{jk}\sigma_{kj},\nonumber\\
\sigma_{jk}&=&\Phi[\langle
\psi_j(\vec{s}_2)|U^{\parallel}|\psi_k(\vec{s}_1)\rangle,
\end{eqnarray}
and $\Phi[z]=z/|z|$. This definition satisfies the requirement of
gauge and reparametrization invariant, hence it is solely a
property of the geometry of Hilbert space and consequently is
measurable. The adiabaticity assumption in \cite{manini} was
subsequently removed \cite{mukunda} and the second order
off-diagonal pure state geometric phase was
verified\cite{hasegawa}. More recently, the study on the pure
state off-diagonal geometric phase has been extended to mixed
quantal states and an experiment to test the off-diagonal mixed
geometric phase was proposed \cite {filipp}. All these studies are
available for single particle systems or composite systems without
intersubsystem couplings.

In this paper, we investigate the effect due to the
inter-subsystem coupling on the off-diagonal phase in composite
systems. This question arises  when we examine the application of
geometric phase in quantum information processing
\cite{zanardi1,jones,falci,duan,wang}, there all systems are
composite and most subsystems interact with each other in order to
store information and implement quantum logic gate. Besides,
entanglement might be created among subsystems via interaction and
it was proved to be a dominant  factor in mixed state geometric
phase \cite{sjoqvist1}. Thus how inter-subsystem coupling may
change the geometric phase of the system is of interest on its
own. We will examine the off-diagonal geometric phase for a
bipartite system consisting of two-coupled spin-$\frac 1 2 $ (or
quantum bit), both of them are driven by time-dependent magnetic
fields. The Hamiltonian describing such a system
reads\cite{abragam} (with $\hbar=1$, hereafter),
\begin{equation}
H(t)=4\xi s_1^z\otimes
s_2^z+\mu\vec{B}(t)\cdot(\vec{s}_1+\vec{s}_2), \label{ha}
\end{equation}
where $\xi$ is the exchange interaction constant (assumed positive
without loss of generality), $\mu$ is the gyromagnetic ratio,
$\vec{s}_j=(s_j^x,s_j^y,s_j^z)$ is the $k$th spin operator
($j=1,2$) composed of the pauli matrices.
$\vec{B}(t)=[B_x(t),B_y(t),B_z(t)]$ represents the time-dependent
magnetic field $\vec{B}(t)$, and we will make use of notation
$\vec{\beta}=(\beta_x,\beta_y,\beta_z)=\mu \vec{B}$ in latter
discussions. The instantaneous eigenstates of $H(t)$ can be
written as
\begin{widetext}
\begin{equation}
|\phi_j(t)\rangle=\frac{1}{\sqrt{N_j}}[
\frac{\beta_x+i\beta_y}{\sqrt{2}}|\downarrow\downarrow\rangle
+(E_j-\xi+\beta_z)|\psi^+_{\uparrow\downarrow}\rangle-\frac{(\beta_x-i\beta_y)(E_j-\xi+\beta_z)}
{\sqrt{2}(\xi+\beta_z-E_j)}|\uparrow\uparrow\rangle],
\end{equation}
\end{widetext}
and the corresponding eigenvalues satisfy
\begin{eqnarray}
E^3-\xi
E^2-(\beta_x^2+\beta_y^2+\beta_z^2+\xi^2)E\nonumber\\
-[\xi(\beta_z^2-\beta_x^2-\beta_y^2)-\xi^3]=0,
\end{eqnarray}
where
$|\psi_{\downarrow\uparrow}^+\rangle=\frac{1}{\sqrt{2}}(|\downarrow\rangle_1|\uparrow\rangle_2
+|\uparrow\rangle_1|\downarrow\rangle_2)$ with
$|\downarrow\rangle_k$ and  $|\uparrow\rangle_k$ denoting,
respectively, the spin-down ($m=-\frac 1 2 $) and spin-up
($m=\frac 1 2 $) states of the $k$th spin.  This Hamiltonian is of
relevance to NMR experiment where Carbon-13 labelled chloroform in
$d_6$ acetone may be used as the sample. The single $^{13}C$
nucleus and the $^1H$ nucleus play the role of the two spin-$\frac
1 2 $; the spin-spin coupling constant in this case is $4\xi\simeq
(2\pi)214.5 \mbox{Hz}$. The quantum system initially in
$|\psi_k(0)\rangle $ driven by the time-dependent magnetic field
would acquire non-zero off-diagonal geometric phases when it
evolves into the other instantaneous eigenstates
$|\psi_j(t)\rangle$ ($j\neq k$), to see this is exactly the case
in our discussion, we propose to use a magnetic field which
rotates in the $xy$ plane at constant frequency $\omega$,
$\beta_x(t)=\Omega(t)\cos\omega t$,
$\beta_y(t)=\Omega(t)\sin\omega t$ and varies linearly in the $z$
direction, $\beta_z(t)=A t$. As shown in Ref. \cite{unanyan01},
the energies of the bare (diabatic, or $\Omega(t)=0$) states cross
at two distinct times (for $\omega<2\xi$) $t_a=(\omega+2\xi)/A$
and $ t_b=\omega/A,$ this was illustrated in figure 1-(a), there
the energies of the adiabatic states have avoided crossings due to
the coupling of the system to the external driving fields.
\begin{figure}
\includegraphics*[width=0.95\columnwidth,
height=0.8\columnwidth]{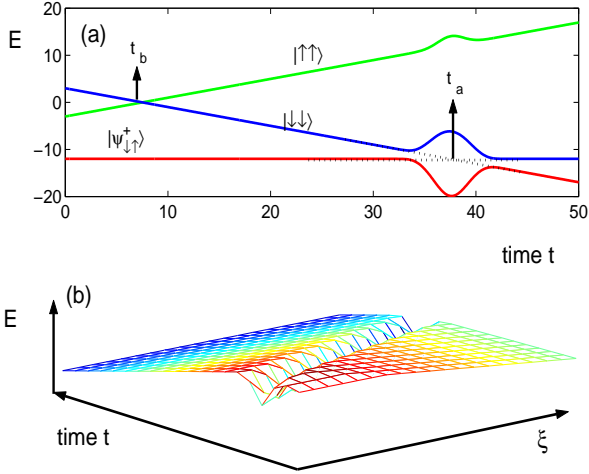} \caption{ (Color
online)Energy(in units of $\xi$) diagram for the two coupled
qubits system. The solid lines in (a) show the instantaneous
eigenvalues of the Hamiltonian Eq.(\ref{ha}), while the dotted
line the diagonal elements of the Hamiltonian. (b) illustrates the
dependence of the lowest eigenenergies on the inter-subsystem
coupling constant. We chose Eq.(\ref{envelop}) as the
transverse-field envelope $\Omega(t)$ and set $\Omega_0=2.5\xi$,
$\omega=0.7\xi$, $A=0.075\xi$ for this plot. The time $t$ was
chosen in units of $1/(2\pi\xi)$} \label{fig1}
\end{figure}
Clearly, state $|\downarrow\downarrow\rangle$ may undergo an
adiabatic evolution ending in
$|\psi_{\downarrow\uparrow}^+\rangle$ by properly designing a
pulse-shaped time dependance for the transverse-field envelope
$\Omega(t)$, with such a choice the field-induced interaction is
maximum at $t_a$ and can be made negligible at the other
crossings. In deed, a Gaussian pulse, centered at $t_a$
\begin{equation}
\Omega(t)=\Omega_0e^{-(t-t_a)^2/T^2} \label{envelop}
\end{equation}
is one of the choices that satisfies the requirements, and the
solid lines in figure 1-(a) are plotted for the instantaneous
eigenvalues  with this choice. The time $t_a$  at which the
adiabatic states have avoided crossings depends on the
inter-subsystem coupling $\xi$ as figure 1-(b) shows, and the
energy separation between those eigenstates at $t_a$ also depends
on $\xi$, the stronger the coupling, the smaller the separation.
So, the inter-subsystem coupling would affect the adiabaticity of
the system. Mathematically, the adiabatic condition can be given
by\cite{allen}
\begin{equation}
\frac{1}{\sqrt{2}}|\Omega \dot{\Delta}-\dot{\Omega}\Delta|<<
(2\Omega^2+\Delta^2)^{\frac 3 2} \label{adiacon}
\end{equation}
with $\Delta=2\xi+\omega-At$.

To proceed further, we introduce new notations
$|1\rangle\equiv|\downarrow\downarrow\rangle, $ $|2\rangle\equiv
|\psi_{\downarrow\uparrow}^+\rangle$ and $|3\rangle\equiv
|\uparrow\uparrow\rangle$ to simplify the representation. With
these notations, the off-diagonal geometric phase factor can be
expressed as $\gamma_{12}(t)=\sigma_{12}(t)\sigma_{21}(t)$ with
\begin{eqnarray}
\sigma_{12}(t)&=&\Phi[\langle 1|\phi_2(t)\rangle],\nonumber\\
\sigma_{21}(t)&=&\Phi[\langle 2|\phi_1(t)\rangle].
\end{eqnarray}
\begin{figure}
\includegraphics*[width=0.95\columnwidth,
height=0.8\columnwidth]{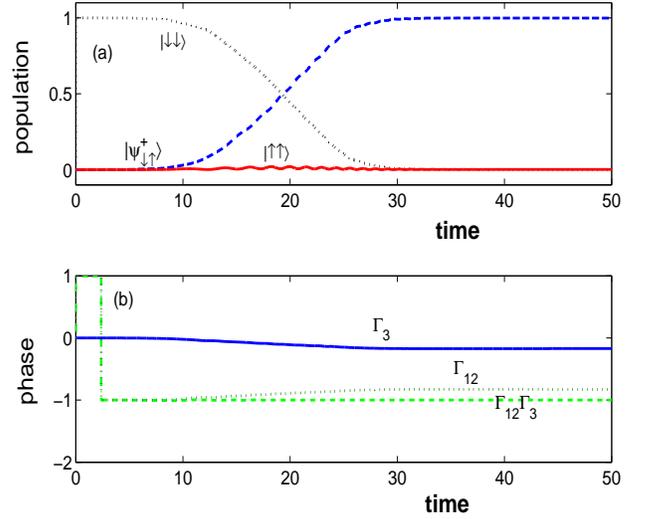} \caption{ (Color online)
Numerical results for the time evolution of the off-diagonal
geometric phase $\Gamma_{12}$( dotted line) and the (diagonal)
geometric phase $\Gamma_3$ acquired in state
$|\uparrow\uparrow\rangle$ as well as the product of
$\Gamma_{12}=arg(\gamma_{12})$ and $\Gamma_3=arg(\gamma_3)$(dashed
line). We chose $|\downarrow\downarrow\rangle$ as the initial
state, and $\Omega_0=0.8\xi$, $\omega=1\xi$, $A=0.16\xi$ for this
plot. The corresponding time evolution of the population with the
same parameters was presented in (a), and the phase was shown in
units of $\pi$, as well as time in units of $1/(2\pi\xi)$.}
\label{fig2}
\end{figure}
Here $|\phi_j(t)\rangle=U^{||}|j\rangle (j=1,2)$ denotes the
parallel transported state starting from $|j\rangle$, for an
adiabatic evolution it would coincide with the instantaneous
eigenstate of  the Hamiltonian except a dynamical phase factor
difference.  In the qubit (two-level) case, it is proved that the
off-diagonal geometric phase factor $\gamma_{ij}$ becomes $-1$ for
any path on the Bloch sphere. The situation under consideration is
quite different from the qubit case, it represents coupled two
qubits driven by a varying magnetic field, since the singlet state
$|\psi_{\downarrow\uparrow}^-\rangle=\frac{1}{\sqrt{2}}(|\downarrow\rangle_1|\uparrow\rangle_2
-|\uparrow\rangle_1|\downarrow\rangle_2)$ is isolated from the
triplet in this model,  it also describes a three-level system
precessing in a magnetic field for the special inter-subsystem
coupling $s_1^z\otimes s_2^z$. The numerical results for the
off-diagonal geometric phase $\Gamma_{12}=arg(\gamma_{12})$ were
shown in figure 2, where we plotted $\Gamma_{12}$ as a function of
time (figure 2-(b), dotted line), the diagonal geometric phase
$\Gamma_3=arg(\gamma_3)=arg(\Phi[\langle\uparrow\uparrow
|\phi_3(t)\rangle])$ was also illustrated for making a contrast
with $\Gamma_{12}$. It is clear that, the system starts acquiring
off-diagonal geometric phases from   $t\simeq
10[\frac{1}{2\pi\xi}]$, when the population begin changing
dramatically, and stops gaining it at $t\simeq
30[\frac{1}{2\pi\xi}]$, from that instant of time the populations
of the three involved levels remain constant. As figure 2-(a)
shows, the two eigenstates, $|1\rangle$ and $|2\rangle$, at the
final point of the path are a permutation of the initial states,
i.e., $|1(\vec{s}_2)\rangle=|2(\vec{s}_1)\rangle$,
$|2(\vec{s}_2)\rangle=|1(\vec{s}_1)\rangle$, where $\vec{s}_1$ and
$\vec{s}_2$ denote the initial and the final point on the path,
respectively. This permutation properties lead to
$\Gamma_{12}\Gamma_3=-1$ (figure 3-(b)) as predicted in
\cite{manini}.

Now, we turn to discuss the effect due to the inter-subsystem
coupling on the off-diagonal geometric phase of the composite
system, the inter-subsystem coupling is fixed for a specific
sample in general, for instance, in NMR experiment the coupling
constant $\xi \simeq (2\pi)214.5Hz$, where $^{13}C$ and $^1H$ in
$d_6$ acetone act as the two qubits. We might change the ratio of
the inter-subsystem coupling $\xi$ to the external magnetic field
driving $|\vec{B}|$ via adjustable quantities $A$ and $\Omega_0$
in Eq.(\ref{envelop}) and $\beta_z(t)=At$, this way we could get
the effect of the inter-subsystem coupling on the off-diagonal
geometric phase. Figure 3 shows the numerical results for
$\Gamma_{12}$ and $\Gamma_{12}\Gamma_3$ as a function of
$\Omega_0$ and $A$.
\begin{figure}
\includegraphics*[width=0.95\columnwidth,
height=0.4\columnwidth]{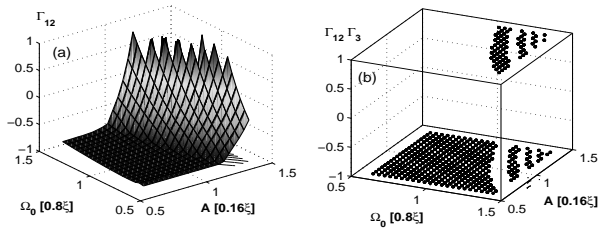} \caption{ (Color online)
The off-diagonal geometric phase (a) and the product
$\Gamma_{12}\Gamma_3$ (b) in units of $\pi$ versus $\Omega_0$ and
$A$. The parameter chosen is $\omega=\xi.$ } \label{fig3}
\end{figure}
For the system to undergo an adiabatic evolution,
Eq.(\ref{adiacon}) must be satisfied, it would make constraints on
$A$ and $\Omega_0$ when we choose Eq.(\ref{envelop}) as the
transverse-field envelop. An alternative constraint  on $A$ and
$\Omega_0$ to ensure the adiabatic evolution is whether
$\Gamma_{12}\Gamma_3=-\pi,$ as it must be satisfied if the system
remains in one of its instantaneous eigenstates in the evolution.
The regime within which the system undergoes adiabatic evolutions
was illustrated in figure 3-(b), while figure 3-(a) was plotted
for $\Gamma_{12}$ {\it versus} $A$ and $\Omega_0$. It is clear
that $\Gamma_{12}$ decreases with $\Omega_0$ and $A$ increasing;
large $A$ leads to large slope in the eigenenergies and then
results in the wide  energy spacing between them. $\Omega_0$
characterizes the transition frequency among basis
$\{|\uparrow\uparrow\rangle, |\psi_{\downarrow\uparrow}^+\rangle,
|\downarrow\downarrow\rangle\}$, consequently describes the energy
gap between the two eigenenergies at time $t_a$. The off-diagonal
geometric phase would depend upon(be proportional to) the
transition frequency among the involved energy levels, the
transition frequency would decrease with the energy spacing
increasing, thus the off-diagonal geometric phase decrease with
$A$ and $\Omega_0$ increasing as illustrated in figure 3-(a).

When there are more than two orthogonal eigenstates are involved
in the permutations, the off-diagonal geometric phase depends on
the decomposition of the permutation \cite{manini}. For instance,
a permutation $P=\left (\matrix{1& 2& 3&\cr 3& 1 & 2} \right )$
can be decomposed as $ 1\rightarrow 3, 3\rightarrow 2,
2\rightarrow 1$. In this case, the off-diagonal geometric phase
factor was defined as
$\gamma_{321}=\sigma_{32}\sigma_{21}\sigma_{13}$, and it would
take value of 1 as proved in \cite{manini}. In our model, we may
realize this process via designing  the transverse-field envelope
$\Omega(t)$. In fact, a twin Gaussian pulse centered at $t_a$ and
$t_b$
\begin{equation}
\Omega(t)=\Omega_{0a}e^{-(t-t_a)^2/T_a^2}+\Omega_{0b}e^{-(t-t_b)^2/T_b^2}
\label{envelop1}
\end{equation}
\begin{figure}
\includegraphics*[width=0.95\columnwidth,
height=0.6\columnwidth]{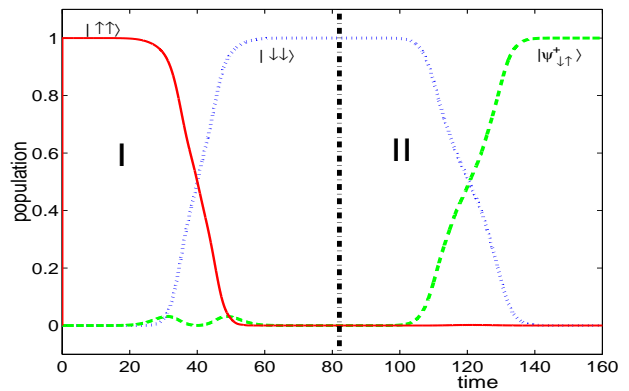} \caption{ (Color
online)Numerical results for the time evolution(time in units of
$1/(2\pi\xi)$  of the populations of the initial state
$|\uparrow\uparrow\rangle$, and state
$|\downarrow\downarrow\rangle$ as well as
$|\psi_{\downarrow\uparrow}^+\rangle$. The parameters chosen are $
A=0.025\xi, T_a=T_b=0.68[\frac{1}{2\pi\xi}], \Omega_{0a}=0.3\xi,
\Omega_{0b}=\xi, \omega=\xi.$ } \label{fig4}
\end{figure}
satisfies the requirement and would lead to the three-eigenstate
involved  permutation. Figure 4 shows the the populations on
states $|\uparrow\uparrow\rangle,
|\psi_{\downarrow\uparrow}^+\rangle$ and
$|\downarrow\downarrow\rangle$ with the system in
$|\uparrow\uparrow\rangle$ initially. It is clear that the
permutation among the involved three orthogonal eigenstates is
completed with the properly chosen parameters.

Remarks and conclusion are in order. In this paper, we studied the
off-diagonal geometric phase in the composite system with
inter-subsystem couplings. Two cases are considered here (a)two
states involved permutation and (b)three states involved
permutations. The latter case yields $+1$ for the off-diagonal
geometric phase factor while the off-diagonal geometric phase
factors depend on the inter-subsystem coupling dramatically in the
former case. These couplings usually can generated entanglement
among the subsystems, then prior shared entanglement, as the
couplings did, would affect the off-diagonal geometric phase of
the composite system. For subsystems that compose a system with
inter-subsystem coupling, there is no effective Hamiltonian
available in general, so the generalization of the pure state
off-diagonal geometric phase to the case of mixed
states\cite{filipp}  is not available for this case, it calls for
further investigations.

\vskip 0.3 cm XXY acknowledges enlightening discussions with Dr.
Nicola Manini, this work was supported by EYTP of M.O.E, and NSF
of China Project No. 10305002.

\end{document}